Hindawi

## Review Article
# The Integral Field View of the Orion Nebula

## Adal Mesa-Delgado

*Instituto de Astrofísica, Facultad de Física, Pontificia Universidad Católica de Chile, Avenida Vicuña Mackenna 4860, Macul, 782-0436, Santiago, Chile*

Correspondence should be addressed to Adal Mesa-Delgado; amesad@astro.puc.cl

This paper reviews the major advances achieved in the Orion Nebula through the use of integral field spectroscopy (IFS). Since the early work of Vasconcelos and collaborators in 2005, this technique has facilitated the investigation of global properties of the nebula and its morphology, providing new clues to better constrain its 3D structure. IFS has led to the discovery of shock-heated zones at the leading working surfaces of prominent Herbig-Haro objects as well as the first attempt to determine the chemical composition of Orion protoplanetary disks, also known as *proplyds*. The analysis of these morphologies using IFS has given us new insights into the abundance discrepancy problem, a long-standing and unresolved issue that casts doubt on the reliability of current methods used for the determination of metallicities in the universe from the analysis of H II regions. Results imply that high-density clumps and high-velocity flows may play an active role in the production of such discrepancies. Future investigations based on the large-scale IFS mosaic of Orion will be very valuable for exploring how the integrated effect of small-scale structures may have impact at larger scales in the framework of star-forming regions.

## 1. Introduction

H II regions are huge volumes of gas associated with recent star formation. These gaseous clouds are mainly ionized and heated by the stellar ultraviolet (UV) radiation emitted by nearby OB-type stars. The study of the elemental abundances of H II regions is an essential tool for our knowledge of the chemical composition and evolution of the Universe, from the local interstellar medium (ISM) to high-redshift galaxies. We still naively tend to describe them as homogeneous Strömgren spheres, but this idealized picture is rarely observed. Instead, reality turns out to be much more complicated, and such regions are found to be highly structured at all scales with complex internal motions. Their morphology results from the structure of the parent molecular cloud, which is affected over time by the UV radiation, stellar winds, or high-velocity ejections associated with star-formation phenomena.

The Orion Nebula (M42, NGC1976) is one of the best studied objects in the sky and the best studied H II region of our Galaxy. The combination of being the nearest H II region, associated with a young stellar cluster containing massive OB-type stars, and having an apparent high surface brightness makes this landmark object a fundamental reference of the solar neighborhood and the subject of multiple investigations by state-of-the-art instrumentation at both ground-based and space observatories. There are outstanding papers in the literature that review in great detail our present knowledge of this object. In particular, for a comprehensive view of the nebula and its associated stellar population, readers are referred to the compilation by O'Dell [1] and the references therein.

Although the Orion Nebula has a large extent of more than one-half degree diameter on the sky, most of the radiation comes from the inner part, the so-called Huygens region (Figure 1). This is an active and complex star-forming region, ionized by a group of four massive stars known as the Trapezium cluster. The main ionizing source of this cluster is named $\theta^1$Ori C (O7V [2]), which is responsible of the bulk of the nebular emission from the main ionization front (MIF). Physical, chemical, kinematical, and structural properties of the MIF have been studied by many authors making use of different observational techniques from X-rays to radio wavelengths (e.g., [3–13]) as well as by means of modeling (e.g., [14–19]).

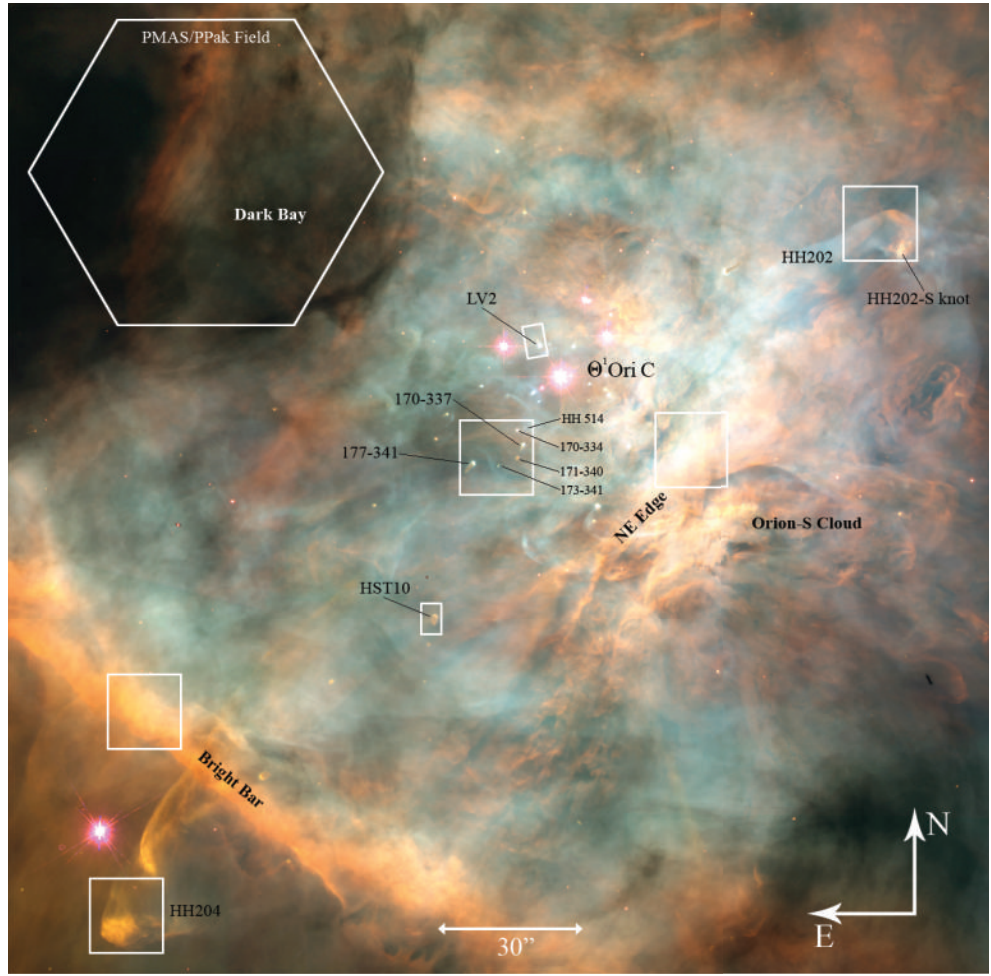

FIGURE 1: Color-composite image of the Huygens region observed with the Wide Field Planetary Camera 2 (WFPC2) on board the Hubble Space Telescope (HST) [6]. The fields of the different IFS studies carried out so far are shown on the image as well as the morphological structures that they contain. Additionally, the location of $\theta^1$Ori C, the Orion-S Cloud, and the Dark Bay are also marked. Squares represent the PMAS studies with the $16'' \times 16''$ integral field unit (IFU) that provides a spatial resolution of $1'' \times 1''$ [37, 53, 54, 65]. Rectangles show the FLAMES studies using the $6.8'' \times 4.3''$ Argus IFU with a spatial resolution of $0.13'' \times 0.31''$ [51, 66, 67]. The early work of Vasconcelos and collaborators [27] with GMOS on the LV2 proplyd had a slightly smaller field of view and spatial resolution than FLAMES. The field of view of the fiber-based PMAS/PPak mode of about 1 arcmin$^2$ is presented on the Dark Bay. As it is seen in **Figure 8**, this IFS mode was used to map the whole Huygens region with a spatial resolution of $2.7''$ [33, 36].

A major characteristic of the Orion Nebula and the Huygens region is that we actually see in great detail the processes related to the ongoing star-formation. The region is well populated with high-velocity outflows, produced by highly energetic ejections of material powered by young pre-main-sequence stars or their possible interactions. Examples of these phenomena are the Herbig-Haro (HH) objects detected in optical studies (e.g., [20–22]) or the molecular outflows observed in the BN/KL and Orion South (Orion-S) regions (e.g., [23, 24]). The Huygens region is also well known for containing hundreds of *proplyds* [25]. This term depicts a special class of protoplanetary disk, resulting from the evolution of a circumstellar disk in the presence of ionizing radiation from massive OB-type stars, and was coined by O'Dell and collaborators [26] to describe the silhouette and tear-drop shaped objects observed in the first imaging studies of the Orion Nebula using the Hubble Space Telescope (HST).

The aim of the present review is to summarize the recent advances that have been achieved in the Orion Nebula through the use of the integral field spectroscopy (IFS). The application of this technique is relatively new in the study of the Orion Nebula and its morphological structures. Indeed, the first work applying IFS in Orion goes back to 2005 [27], where the properties of the famous LV2 proplyd [28] and its associated outflow were studied. Since then, a total of 8 new studies have come out in the literature using this observational technique. Readers are referred to the original works for a detailed description of the reduction and analysis methods. The fields that have been mapped with IFS in these studies are shown in **Figure 1**. It is noted that up to today IFS has been used to investigate physical, chemical and structural properties of the MIF, HH objects, and proplyds. In the following sections, an extensive revision of these works is presented.

The Orion Nebula is also an excellent testing lab to investigate the abundance discrepancy (AD) problem, one of the major still unresolved issues in the physics of photoionized nebulae (see [29]). This problem arises from the observational fact that chemical abundances of heavy-element ions determined from the bright collisionally excited lines (CELs) are systematically lower than the abundances derived from the faint recombination lines (RLs) emitted by the same ions. IFS studies in the Orion Nebula have found new and interesting clues in the context of this problem, which are also reviewed here.

## 2. Resolving the Ionization Structure

One of the major capabilities of the IFS is the access to the bidimensional information, allowing us to spatially resolve the ionization structure of the ionized gas in Galactic and nearby extragalactic objects. Studying the ionization structure of the Orion MIF and its physical and chemical properties is a topic that has been addressed by several authors following different observational approaches: echelle and long-slit spectrophotometry (e.g., [9, 11, 15, 30]), Fabry-Perot imaging spectrophotometry [31], and CCD imagining (e.g., [5, 32]). However, it was not until 2007 that this topic was studied in Orion making use of IFS.

Sánchez and collaborators [33] were the first to conduct a global analysis of the Huygens region through IFS. Centered around the Trapezium cluster, a big mosaic was made by these authors from observations with the Potsdam Multi-Aperture Spectrograph (PMAS [34]) at the 3.5 m telescope of Calar Alto Observatory (Almería, Spain). The observations used the fiber-based integral field unit (IFU) of PMAS known as PPak [35], which provides a hexagonal field of view of about 1 arcmin² on the sky and a spatial resolution of 2.7″ (see Figure 1). Thanks to this dataset, it was possible to analyze the integrated properties of the whole nebula for the first time from spectroscopic observations. The physical conditions derived from this analysis were in general consistent with the results of previous spectroscopic works. An important amount of continuum emission in the blue was observed in the integrated spectra, which might be associated with diffuse continuum emission of scattered light (e.g., [35]). Unfortunately, the spectral and spatial resolution of this dataset were not high enough to address this topic. The spatial distribution maps of emission line fluxes, dust extinction, electron density and temperatures, and the He and O abundances were obtained by these authors and allowed them to detect the effects of well-known morphologies in Orion such as the Bright Bar, the Dark Bay, or prominent HH objects. A relation between the He abundance and the ionization structure that was interpreted as possible deviations of the case B recombination theory was also found. As it will be described in Section 6, this dataset presented certain limitations that have been recently considered, and an improved mosaic of Orion has been obtained with the aim of investigating the previous issues with the required accuracy (see Section 6 and preliminary results in [36]).

Analyzing the MIF is of special interest since it is the transition layer between the fully ionized gas and the neutral surroundings, where density and temperature gradients are expected. Though the MIF is found in almost every direction towards the Huygens region, only certain geometries allow us to explore its ionization structure in depth. The author of this review and collaborators [37] presented the first dataset devoted to study the ionization structure of the MIF applying IFS on two conspicuous features of Orion: the Bright Bar (BB) and the northeast edge of the Orion-S cloud (NE-Orion-S). The observations were performed with PMAS and the $16″ \times 16″$ IFU. With this setup, emission line fluxes and ratios, physical conditions, and chemical abundances were mapped at spatial scales of $1″ \times 1″$. The location of the fields can be found in Figure 1.

The spatial distribution of the [O I]/Hα, [N II]/Hα, and [O III]/Hβ line ratios in both BB and NE-Orion-S fields are plotted in Figure 2 for illustration. The maps clearly show the ionization stratification in the two morphological structures, which is the combined result of different inclination angles along the line of sight and different distances to $\theta^1$Ori C. On the one hand, the BB is located at about $111″$ to the southeast of $\theta^1$Ori C, seen as an elongated structure in both ionized gas (e.g., [20]) and molecular emission (e.g. [38]). In the current picture, the BB is viewed as an escarpment with an average inclination of 7° with respect to the line of sight, where the MIF changes from a face-on to an edge-on geometry [18, 39]. The fact that the MIF is almost edge-on allows us to nicely resolve the ionization structure of the BB as in Figure 2. On the other hand, the NE-Orion-S edge has the peculiarity of being the brightest zone of the Huygens region (see Figure 1), located about $30″$ to the southwest of $\theta^1$Ori C. Recently, the analysis of O'Dell and collaborators [12] suggested that Orion-S is a cloud suspended within the main body of the Orion Nebula in front of the MIF, more tilted than the BB, and ionized only on the side facing to $\theta^1$Ori C. Comparing the ionization structure resolved by PMAS of the BB and NE-Orion-S, it was possible to estimate that the main plane that may contain the NE-Orion-S edge is tilted in average about $48° \pm 13°$ with respect to the line of sight [37].

The PMAS results also showed that the NE-Orion-S is much denser than the BB in consistency with previous studies [31, 33]. Gradients in the electron density, $n_e$, were observed ranging from 3000 to 5000 cm$^{-3}$ in the BB and from 8000 to 16,000 cm$^{-3}$ in the NE-Orion-S. The structure of the electron temperature, $T_e$, turned out rather featureless with relatively low gradients toward the direction of $\theta^1$Ori C. The $T_e$([N II]) maps presented variations from 9000 to 9600 K in the BB and from 9000 to 10,500 K in the NE-Orion-S, while the $T_e$([O III]) ranges from 8200 to 8800 K in the BB and from 8300 to 8800 K in the NE-Orion-S. In the two fields, the maps of total O abundance—as derived from the sum of O$^+$ and O$^{2+}$ CEL abundances—showed mean values consistent with previous determinations of the literature for the Orion Nebula (~8.50 dex [9]). However, rather than being homogeneous as we would expect, they exhibit a structure very similar to the spatial distribution maps of the O$^+$/H$^+$ ratio and $n_e$ ([S II]). The most plausible explanation is that collisional de-excitation would be affecting the [S II] 6717, 6731 Å emission lines because the reported densities

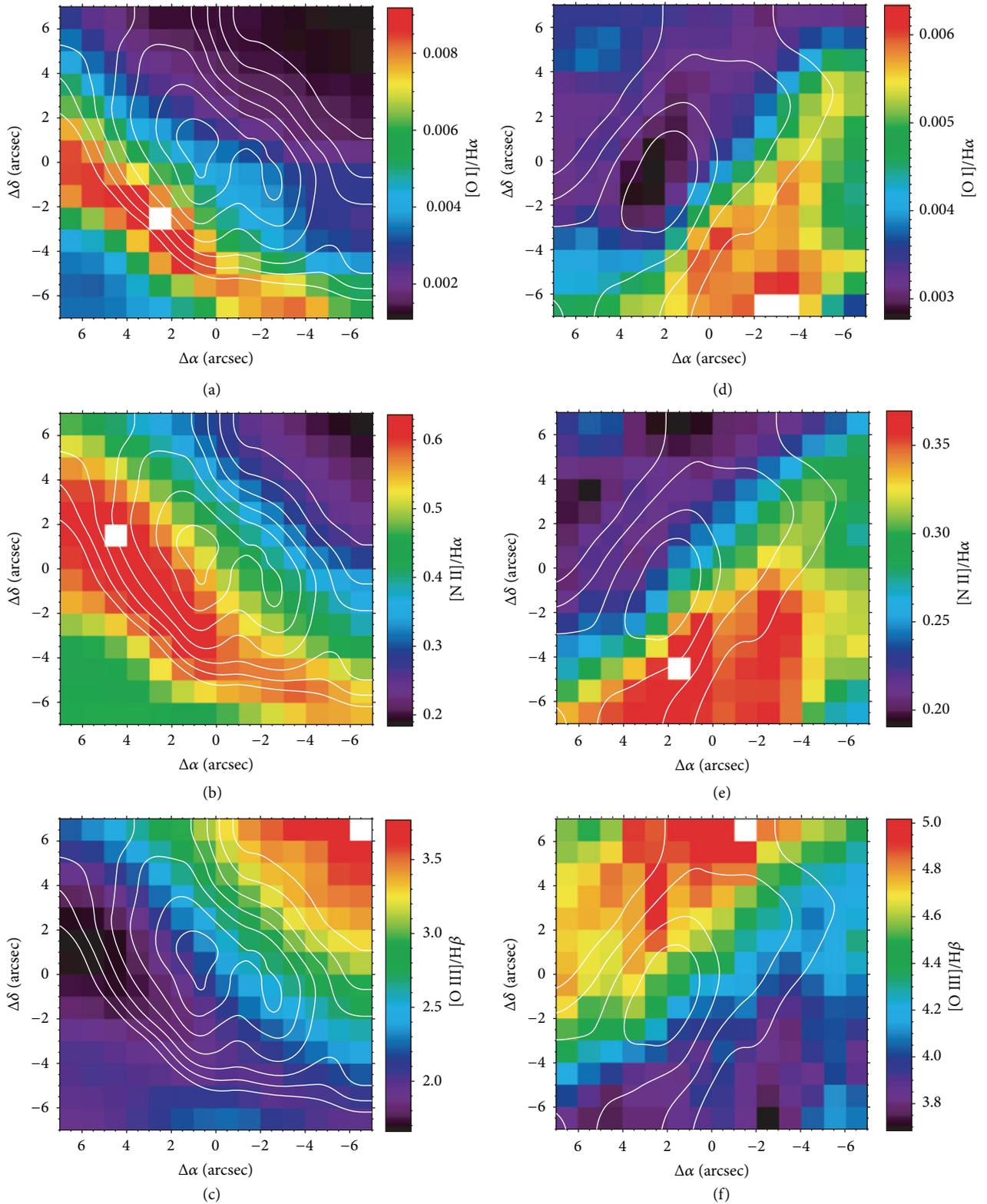

FIGURE 2: Spatial distribution maps of selected emission line ratios for the Bright Bar (left-hand side) and NE-Orion-S (right-hand side) taken from [37]. The sum of [N II] 6548, 6584 Å and [O III] 4959, 5007 Å nebular lines have been used to derive the [N II]/Hα (middle) and [O III]/Hβ (bottom) ratios, respectively. Hα contours are overplotted in all maps.

are—especially in the case of the NE-Orion-S—close to their critical densities ($\sim$3000 cm$^{-3}$ for $\lambda$6717 and $\sim$10,000 cm$^{-3}$ for $\lambda$6731). This effect would lead to biased density estimates, which then would affect the chemical abundance calculations, in particular of those ions that are more dependent on the adopted physical conditions, as for instance O$^+$ (see [37]).

## 3. Effects of High-Velocity Outflows

In last decades, and especially after the launch of the HST, many outflows such as HH objects or collimated jets have been identified in the central part of the Orion Nebula (see e.g., [40–42]). CCD imagining observations (see e.g. [40]) revealed that several of these high-velocity flows are mainly dominated by photoionization from $\theta^1$Ori C, though the possibility cannot be completely discarded that a partial contribution of shocked gas might still exist. It is important to understand the relevance of this contribution and how it affects to the traditional methods used to determine physical and chemical properties in photoionized gas. In the literature, most spectroscopic studies of Orion HH flows have been focused on studying their gas kinematics (e.g., [21, 22]), while there is an important lack of detailed analysis of their physical and chemical properties as well as their effects on the surrounding media. Until today, this subject has been mainly addressed theoretically (e.g., [43, 44]); and sparsely investigated observationally only making use of high-resolution echelle spectroscopy in a few objects: HH529 [45], the south knot of HH202 (HH202-S [46]), HH888, and HH505 [47], and the microjet arising from the LV2 proplyd [48].

The incorporation of the IFS in the analysis of gas flows in Orion have been investigated since few years ago. Vasconcelos and collaborators [27] were the first using IFS in the Huygens region, focusing on the LV2 proplyd (see Figure 1) and, mainly, on its microjet. Intermediate-resolution ($R \approx 5500 \approx 55$ km s$^{-1}$) observations were performed in the spectral range 5500–7500 Å with the Gemini Multiobject Spectrograph IFU (GMOS [49]) at the Gemini South Observatory. Though observations were limited by the seeing, the great spatial resolution of the GMOS IFU, $\sim$0.2$''$, allowed the authors to spatially resolve the red-shifted jet from the proplyd. The authors also found traces that pointed out to the presence of a blue-shifted component of the jet, previously reported by Doi and collaborators [50]. From the H$\alpha$ flux in the jet, the mass-loss rate of the proplyd was estimated in this work, being consistent with previous estimations [48].

Recently, Tsamis and Walsh [51] have presented an improved IFS dataset with respect to those of Vasconcelos and collaborators. High-resolution ($R \approx 30,000 \approx 10$ km s$^{-1}$) observations were accomplished with the Fibre Large Array Multi-Element Spectrograph (FLAMES; [52]) and the Argus 6.8$''$ × 4.3$''$ IFU mode at the 8.2 m *Kueyen* Very Large Telescope (VLT). The high-quality of the dataset allowed these authors to characterize for the first time physical and chemical properties of the bipolar jet arising from LV2. Though the original spatial resolution provided by the selected Argus mode (0.31$''$×0.31$''$) was limited by the seeing of the observing night ($\sim$0.8$''$), both red- and blue-shifted lobes were marginally spatially detected in several emission

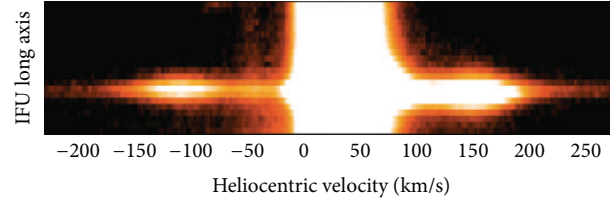

FIGURE 3: Image adapted from [51] showing the H$\alpha$ emission profile of the LV2 proplyd observed with FLAMES with an effective velocity resolution of about 2.3 km s$^{-1}$ pixel$^{-1}$ (FWHM velocity resolution of 7.7 km s$^{-1}$). The red- and blue-shifted components of the bipolar jet are clearly resolved, even without subtracting the bright background emission of the nebula. The intensity scale is logarithmic with a minimum of $5.4 \times 10^{-17}$ (black) and a maximum of $1.1 \times 10^{-13}$ ergs$^{-1}$ cm$^{-2}$ (white). The length of the vertical axis is 6.6$''$.

lines. Spectrally, the jet emission was well resolved as we can see in Figure 3.

In the work of Tsamis and Walsh, an accurate study of the surface brightness distributions of the red- and blue-shifted lobes in H$\alpha$, H$\gamma$, and [Fe III] 4658 Å led the authors to conclude that the bipolar jet is being ejected in a projected axis almost perpendicular to the tail of the LV2 proplyd. From this result, it is clear that the use of IFS combined with high spectral resolution is a refined technique that can reveal unexpected properties that, on the other hand, would remain hidden—or would be more difficult to identify—under the view of the classical slit spectroscopy. Another interesting result is the drastic temperature variations observed in the velocity profile of the $T_e$([O III]), associated with the red-shifted emission of the jet, which might be produced by a shock discontinuity as the authors argued. Finally, a noteworthy result of this work is also the significantly enhanced Fe abundance in the jet emission, pointing to very efficient dust destruction mechanisms that might be operating in high-velocity irradiated jets as was also reported in the study of the HH202-S knot [46].

Models of photoionized HH flows [43] predict that the $T_e$ along an HH object is of about the typical value of a gas in photoionization equilibrium—about $10^4$ K—but shows a localized increase at the leading working surface of the bow shock due to shock heating. This zone is narrow and precedes the high-density shocked gas behind the working surface of the gas flow. Very recently, the discovery and localization of this structure have been possible thanks to the use of IFS. Making use of PMAS, its 16$''$ × 16$''$ IFU and spatial scales of 1$''$ × 1$''$, the presence of this structure seems to have been found in the form of a high-$T_e$ arc at the leading working surface of the prominent HH204 object of the Orion Nebula [53]. Though it was not reported, a similar narrow arc was also observed in the IFS study of HH202 using the same instrumentation and setup (see [54]). In Figure 4, the detection of these high-$T_e$ arcs in the temperature distribution maps of HH204 and HH202 is illustrated. In both objects, the arcs were detected in the $T_e$ derived from the nebular and auroral [N II] line ratio. According to C. Morisset (private communication), the origin of these arcs should be regarded with caution because they may be artifacts produced by observations of two unresolved gaseous components with very different physical conditions. This clearly points out the need for confirming the true nature of the high-$T_e$ arcs through high-resolution spectroscopy, resolving the kinematical component associated with the gas flow from the background emission. This kind of observations has been already done in the HH202-S knot [46], though unfortunately the slit did not cover the small high-$T_e$ arc detected in this knot from the IFS study of HH202 (Figure 4).

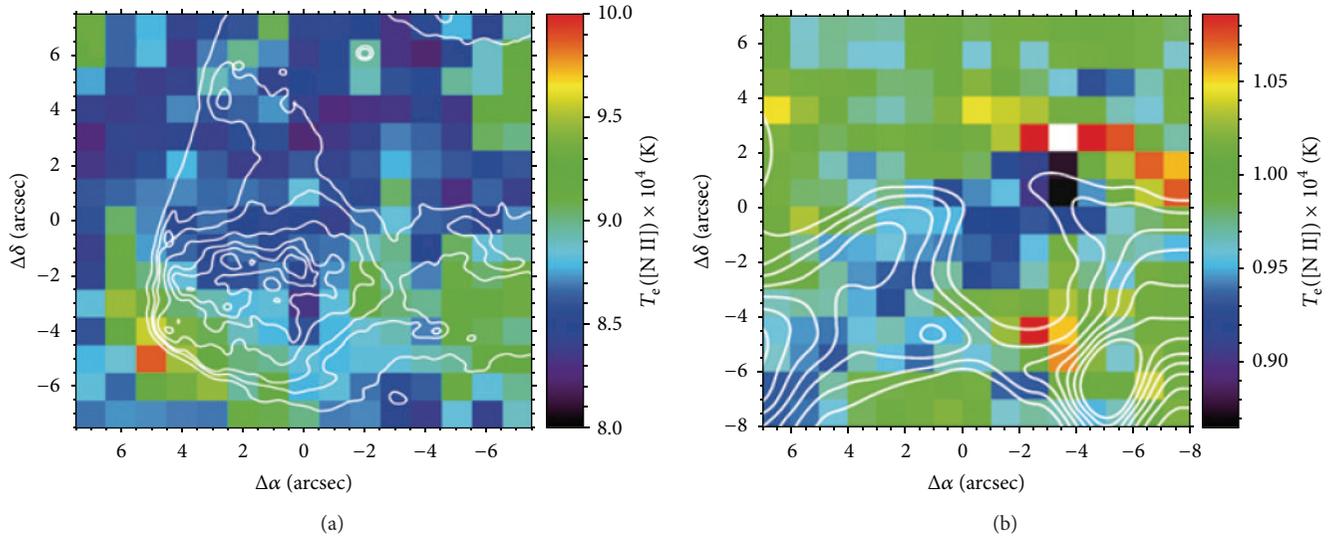

FIGURE 4: Spatial distribution maps of $T_e$([N II]) with Hα contours overplotted for (a) HH204 and (b) HH202, adapted from [53, 54], respectively. High-$T_e$ arcs are located along the narrow areas at the leading working surfaces of the gas flows: at the lower-left corner for HH204 and near the upper-right corner in HH202. An $T_e$ enhancement that might be associated to a small high-$T_e$ arc is also seen near the HH202-S knot.

Núñez-Díaz and collaborators [53] quantified the physical properties of the narrow shock-heated zone found in HH204, measuring a temperature enhancement of about 1000 K with respect to the ambient gas. From their analysis, it is concluded that the compression and heating of the gas due to the presence of high-velocity flows can directly affect the chemical abundance determinations due to: (a) an overestimation of the collisional de-excitation effects on emission lines arising from levels with low critical densities; and (b) the use of too high $T_e$ values for deriving abundances due to contamination from emission of the leading working surface. It is fundamental to investigate in depth the importance of these disturbing effects into the determination of chemical abundances in the Orion Nebula and, in general, H II regions, especially for those objects where these small-spatial scale phenomena cannot be resolved.

## 4. Chemical Composition of Orion Proplyds

Metallicity plays an important role in the evolution of circumstellar disks and in their associated potential to form planets (see [55]). The positive correlation found between the host star metallicity and the presence of giant planetary companions (see, e.g., [56]) has raised great interest in the question of the chemical composition of planet formation circumstellar envelopes. Certainly, Orion proplyds are unique targets to investigate the metallicity content and its role in the evolution of circumstellar disks. The currently accepted model (e.g., [57–59]) is that far-ultraviolet photons heat the surface of the disk, forming a warm outflowing envelope of neutral gas that becomes photoionized after expanding to a few times the outer radius of the disc (see Figure 5). Though they are complex structures, the optical emission of the ionized photoevaporation flow provides an ample variety of emission line features, which can be used as robust diagnostics to determine physical and chemical properties of the gaseous phase. In particular, metallicity can be estimated through the usual proxy, the total O abundance, since O is the most abundant heavy element in the Universe.

The chemical composition of Orion proplyds has, however, remained unknown for a long time. Usually, it has been assumed to be roughly the same than the parent molecular cloud, given by the better known gaseous [9] and stellar [60] abundances of the Orion Nebula. Though spectroscopical studies have been carried out in several Orion proplyds (e.g. [27, 48, 61, 62]) and tentative evidence has been found about the evolution of dust properties in the thick molecular disk [63] and in the ionized photoevaporation flow [64], until very recently there was an important lack of comprehensive studies of the chemical abundances in any proplyd. In the last years, great advances have been done by few authors to correct this situation, where the use of IFS has turned out essential.

The spectroscopical analysis of proplyds is without any doubt a challenging task (see [62, 65]). The major difficulty lies in disentangling the intrinsic emission of the proplyd from possible sources of contamination, especially in low-resolution spectroscopical observations. In principle, the main source of contamination is the emission from the background of the nebula, which has to be properly subtracted. However, it is difficult to define an accurate background since the nebula shows significant brightness variations at small spatial scales. From an observational point of view, a background estimation can be obtained as the average of the nebular emission around the proplyd. In this sense, IFS can be more accurate and effective than slit spectroscopy, giving us more information of the nebular emission surrounding the proplyd. Depending on the particular design of the used spectrograph, IFS has also the advantages of ensuring a major collection of the total flux emitted by the proplyd and the capability of resolving spatially its structure. Instead, slit spectroscopy is more subjected to certain observational effects (selected slit width or seeing changes during the observing night) that could lead to flux losses as well as could complicate the background estimation and subtraction procedures. Independent of the observational strategy, the opacity due to the dust inside the proplyd is another unknown factor that may introduce systematic errors in the subtraction process of the nebular emission. At least an estimation of the proplyd extinction could be provided; physical and chemical properties are limited to two extreme possibilities: (1) a fully opaque case or (2) a fully transparent case. Finally, another possible source of contamination is the presence of high-velocity emission from any jet associated with the proplyd, a problem that can be only solved by high-resolution spectroscopy or focusing on proplyds that do not present such jets (see [42]).

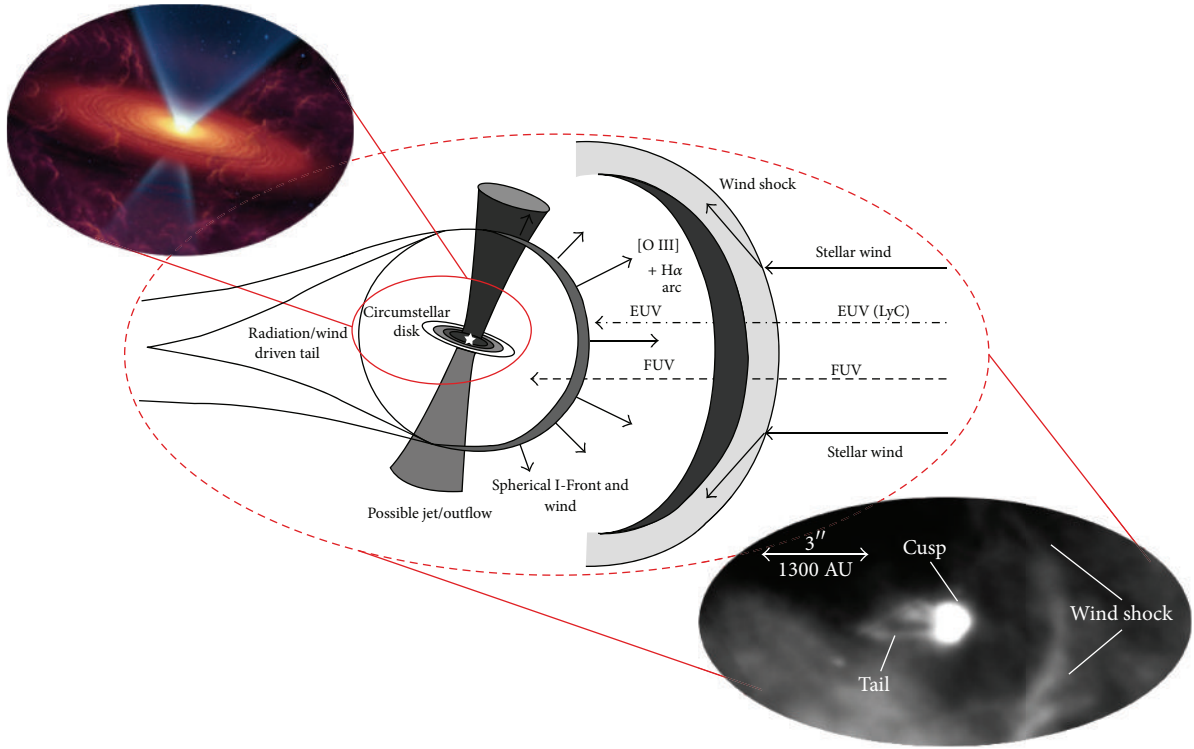

Figure 5: Schematic representation of a proplyd [61], where the young star has a circumstellar disk and a bipolar outflow. The scheme is accompanied by an artistic depiction of the circumstellar disk (upper left-hand side; Subaru Telescope Press Release on 31 August, 2005) and an Hα image of the proplyd HST1 (lower right-hand side [6]). Stellar extreme-UV (EUV) and far-UV (FUV) photons enter from the right. The FUV photons penetrate the surface of the circumstellar disk around the star, driving a slow neutral flow ($\sim$3 km s$^{-1}$), which for most proplyds [57, 58] accelerates to mildly supersonic velocities before shocking and passing through an ionization front (I-Front) at a distance of a few disk radii. In the I-Front the gas is rapidly accelerated to about 10–20 km s$^{-1}$ and continues to accelerate as it expands away from the I-Front and reaches progressively higher stages of ionization due to the EUV photons. The interaction between the photoevaporation flow and the stellar wind can produce a wind shock in front of the proplyd. The neutral flow in the tail is fed by diffuse UV photons, which evaporate the back side of the disk, and possibly also by gas that left the front side of the disk but was redirected into the tail by pressure gradients in the shocked neutral layer. The ionized flow from the tail is induced by diffuse EUV photons, but stellar EUV photons entering from the side also play an important role in maintaining the ionization of the tail flow once it has left the I-Front, especially toward the front of the tail [59].

Currently, under the previous considerations, physical and chemical properties have been so far studied in three Orion proplyds: LV2 (167-317), HST1 (177-341), and HST10 (182-413). Out of this sample, the presence of high-velocity jets has been only reported in LV2 [42]. Intermediate-resolution ($R \approx 12,000$) spectroscopical observations of LV2 and HST10 were performed by Tsamis and collaborators [66, 67] with FLAMES and the setup already mentioned in the observation of the LV2 jet (see Section 3). The proplyd HST1 was observed at low spectral resolution ($R \approx 1500$) in the range 3500–7200 Å with PMAS by the author of this review and collaborators, as part of a larger IFS project in the Orion Nebula that included the PMAS observations already presented in **Section 2** and **Section 3**. As it is seen in Figure 1, the PMAS field containing HST1 (named 177–341 from its coordinates [68]) also includes other features. However, HST1 was the only proplyd with enough signal-to-noise to determine physical conditions and chemical abundances. In **Figure 6**, it is presented a sample of spatial distribution maps in different emission lines to illustrate how these proplyds are seen from IFS.

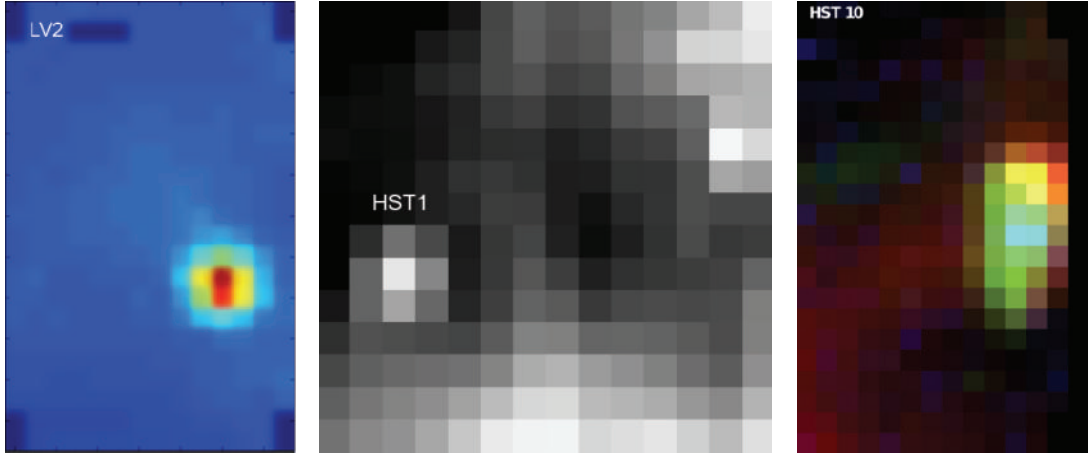

FIGURE 6: Spatial distributions maps of LV2 in Hβ [66], HST1 in Hα [65] and HST10 as a combination of Hα (red), [O I] 6300 Å (green), and [O I] 5577 Å (blue) [67]. A comparison with **Figure 1** is recommended.

A global analysis of the results found in the three proplyds lead us to reach three conclusions that should be considered in future works in relation to the calculations of physical and chemical properties. Firstly, the structures associated with the proplyds and observed in the spatial distribution maps of several nebular properties (e.g., emission line fluxes, line ratios, or electron densities and temperatures) are dominated by collisional deexcitation effects due to the high densities of the proplyds. To properly account for this mechanism, a very detailed analysis of the proplyd spectra is mandatory in order to evaluate the reliability of the density diagnostics (see e.g., [65]). Assuming the full transparent case, electron density values of $2 \times 10^6$, $4 \times 10^5$, and $10^5$ cm$^{-3}$ were found at the cusp of the proplyds LV2, HST1, and HST10. Lower densities were found along the tails of LV2 and HST10. Though a large variation range is noted in the proplyd densities, temperatures derived from the [O III] line ratio turned out rather similar in three cases with values of about 8000–9000 K. Secondly, chemical abundances derived from CELs should preferentially use CELs with high critical densities to minimize collisional deexcitation effects because of the high densities in most proplyd cusps. Low-critical density CELs can be completely suppressed by such effects as it was shown in the analysis of HST1 and LV2. For instance, the O$^+$ abundance is crucial for the determination of the total oxygen abundance and, therefore, it should be calculated from observations of the high-critical density [O II] lines at 7320 + 30 Å (∼ $3 \times 10^6$ cm$^{-3}$) rather than the low-critical density [O II] line at 3727 Å (∼5000 cm$^{-3}$). Finally, the effects associated to the internal extinction by dust in the proplyd's neutral core are appreciable and may severely affect density determinations and ionic abundances that are sensitive to the adopted density. As it was investigated in HST1, density determinations can range from $4 \times 10^5$ cm$^{-3}$ in the full transparent case to $9 \times 10^4$ cm$^{-3}$ in the full opaque case.

In order to investigate the chemical content of proplyds for a better understanding of their metallicity and evolution, intensive collaborations between modelers and observers are encouraged. As an alternative to the empirical analysis, the chemical compositions of proplyds can be also explored constructing physical models for the photoevaporation flows. These models combine simulations of radiative transfer, hydrodynamics, and atomic physics to predict fundamental parameters like the density, temperature, and ionization structure of the photoevaporation flows through the proplyd ionization front (see **Figure 5** [65, 69]). For LV2, the modeling is still in progress (Flores-Fajardo et al. in prep.) and at the moment we count on the empirical abundances based on [O II] and [O III] CELs derived by Tsamis and collaborators [66]. For HST1, total O abundances were only determined from the photoevaporation model due to the absence of [O II] CELs in the intrinsic spectra of the proplyd. On the other hand, HST10 is the only case in which the total O abundance in the gaseous phase was estimated empirically and in the model from the analysis of [O II] and [O III] CELs. The comparison of the O abundances in the three proplyds is rather disparate, so it is difficult to see a consistent trend in their results. Although we find a roughly solar O abundance for HST10 from both empirical analysis and photoevaporation modeling, the O/H ratio was found to be almost twice solar in LV2 from the purely empirical analysis. In contrast, HST1 shows an O abundance of about ∼0.4 × solar from the photoevaporation model fitting. Given the wide range of characteristic ionizations and densities found in the three proplyds, it is possible that systematic errors might be contributing to this abundance spread. To rule out any such effects, it is required to increase our knowledge on the chemical content of proplyds and perform similar analysis on a sample of proplyds with similar physical properties. Though IFS is a unique technique to investigate the metallicity in Orion proplyds, further work is completely necessary before a definitive statement can be made about their gas-phase abundances.

## 5. The AD Problem from IFS

### 5.1. Consequences and Origin.

The AD problem is far from negligible in the analysis of Galactic and extragalactic H II regions. A particular sensitive case is the O/H ratio, which is the most widely used proxy of the global metallicity Z. Observational studies have found that O abundances calculated from the O II 4630–4670 Å multiplet RLs are between 20% and 70% higher than those derived from the [O III] 4363, 4959, and 5007 Å CELs (e.g., [70–75]). Such discrepancies have direct effects on our current knowledge of the chemical composition and chemical evolution in the Universe, affecting (1) the calibration of the strong line methods as the $R_{23}$–O/H Pagel's relation [76, 77], (2) the mass-metallicity and luminosity-metallicity relations [78], (3) the basic ingredients of chemical evolution models and predicted stellar yields [79], (4) the possible metallicity dependence of the Cepheid period-luminosity relation [80], (5) the metallicity dependence of the number ratios of the different types of W-R stars [81], and (6) the determination of the primordial helium [82].

What is the reason of this discrepancy? And which are the emission lines that we should trust? These are the two fundamental questions that after decades of intensive researching remain open and without a satisfactory answer. Traditionally, the AD has been associated with the presence of temperature fluctuations (of still unknown cause) as proposed by Peimbert more than 40 years ago [83–85]. According to this scenario, RLs should provide the correct abundances because their emissivities have a weaker temperature dependence than CELs—more affected by the presence of such fluctuations. From the proposal of Tsamis and Péquignot [86], Stasińska and collaborators explored the hypothesis of inhomogeneous abundances in the ISM (see [87]). They concluded that if this is the real scenario, then the chemical abundances derived from RLs and CELs should be upper and lower limits, respectively, to the true ones, though those from CELs should be more reliable. Very recently, [88] have proposed the possibility that electrons may depart from a Maxwell-Boltzmann equilibrium energy distribution, especially affecting the CEL emission. It is necessary to emphasize that the underlying assumptions of this theory are in contradiction with what has been established for half a century for the conditions in gaseous nebulae. Until the origin of this discrepancy is well understood, chemical abundances based on the standard CEL-method, which is used in the vast majority of cases, especially at extragalactic scales, should be regarded with caution.

### 5.2. Role of Small-Spatial Scale Structures.

Given its proximity, the Orion Nebula is the perfect target to investigate the possible relation between the AD problem and the presence of morphological structures. Tackling this issue certainly requires reliable detections of RLs emitted by heavy-element ions to investigate the RL-CEL discrepancy. From optical long-slit spectroscopy with the Intermediate Dispersion Spectrograph and Imaging System (ISIS) at the 4.2 m William Hershel Telescope, the author of this review and collaborators addressed this topic for the first time in the Orion Nebula at spatial scales of 1.2″ [11]. Very deep observations were performed in five slit positions of 3.7″ long each. The slits were arranged on the Huygens region, covering different morphological structures such as proplyds, HH objects, and stratified bars. A total number of 730 one-dimensional spectra were extracted and reliable detections of the O II multiple 1 RLs were reported in 92% of them. Then, the authors could analyze the spatial distribution profiles of the RL-CEL discrepancy of $O^{2+}$ abundance, which is usually quantified through the AD factor (ADF). In this review, we adopt the logarithmic form of the ADF, defined as the difference of abundances derived from RLs and CELs. One of the major results of this work was that the ADF($O^{2+}$) remains rather constant along most of the observed areas of the nebula but showing localized enhancements at the positions of the prominent HH objects HH202, HH203, and HH204. On average, the ADF($O^{2+}$) is about 0.15 dex, while in the HH areas, the discrepancy increases up to 0.3–0.5 dex.

Incorporating IFS has enormously improved our ability to spatially locate with much more precision areas on the nebula having high AD. This is well illustrated in the IFS analysis of the NE-Orion-S edge [37] and HH202 [54], where it was possible to map the emission of O II RLs in both structures. The ADF($O^{2+}$) maps of these two fields are shown in Figure 7. In the case of NE-Orion-S, the ADF($O^{2+}$) is slightly higher at the north-east corner of the field, though it does not seem to be related to the presence of any remarkable morphology when we compare it with the HST images of the Huygens region at that exact position (see Figure 1). On the contrary, the results found in the ADF($O^{2+}$) map of HH202 are encouraging: the maximum ADF($O^{2+}$) is located at the position where the gas flow reaches its maximum velocity, the HH202-S knot. The same research group carried out a subsequent study of HH202-S in which the emissions from the gas flow and the nebular background were spectrally resolved thanks to the high spectral resolution of the observations ($R \approx 30,000$). Interestingly, the ADF($O^{2+}$) in the gas flow component turned out to be $0.35 \pm 0.05$ dex, a much higher discrepancy than the value of $0.11 \pm 0.04$ dex found in the ambient gas. These results also confirm what was found in the long-slit study and suggest a possible connection between high-velocity flows and high AD. To clearly establish the possible role of high-velocity flows in the AD problem, further investigation is still needed. The use of high-spectral resolution IFS would be the ideal observational approach.

The IFS studies of the proplyds HST1 and LV2 have also brought new clues into the AD problem (see [65, 66]). The possible role of proplyds was also attempted in the early ISIS work presented above, but those observations did not count on reliable diagnostics to properly determine the proplyd densities. A striking result found in HST1 and LV2 is that the ADF($O^{2+}$) tends to zero when physical conditions of proplyds are well accounted as in both full opaque and transparent cases. It is observed that the high densities of proplyds produce a clear enhancement of the $O^{2+}$ abundances derived from CELs with respect to the nebular background abundances, while those form RLs are basically similar in both cases. From these results, it is concluded that high-density gas (in the form of proplyds, globules, or unseen clumps/filaments acting at small spatial scales) may be playing a major role in the AD problem as Viegas and Clegg proposed in 1994 [89]. In a scenario where small, high-density, and semiionized clumps/filaments are mixed with diffuse gas in the observation aperture, the classical method based on CELs can be severely affected if collisional deexcitation of certain emission line diagnostics is not well accounted for. Instead, RLs would not be affected and, therefore, they should reliably yield the chemical abundances in the target field. Judging by the results of the proplyd IFS analyses thus far, these clumps do not need to be strongly hydrogen deficient, unlike those posited in previous scenarios to explain the AD problem in H II regions [86, 87].

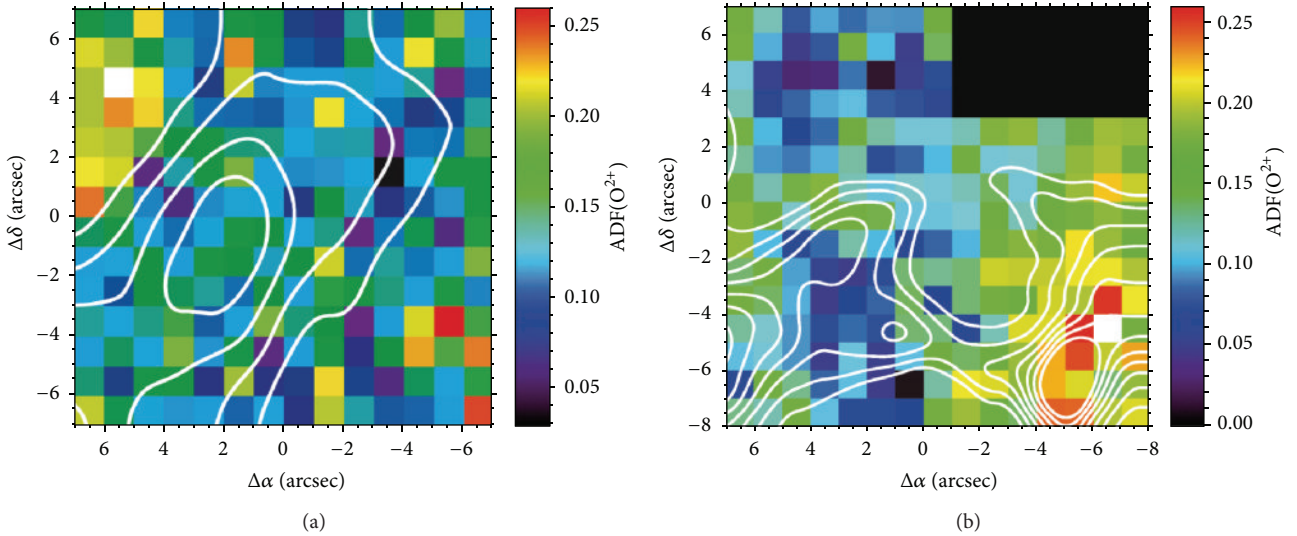

FIGURE 7: Spatial distribution maps of ADF(O$^{2+}$) with H$\alpha$ contours overplotted for (a) NE-Orion-S and (b) HH202, adapted from [37, 54], respectively. The black rectangle in the upper-right corner of HH202 corresponds to a masked area where O II RLs were marginally detected.

## 6. A Deep Global View: The Big Mosaic of Orion

The big mosaic of the Huygens region constructed by Sánchez and collaborators [33] using IFS stands for a step ahead in the application of this technique. However, this dataset needed a significant improvement in at least three aspects. The dataset was poorly flux-calibrated, based on really short exposure time (2 s) and low spectral resolution. Today, these aspects have been improved and a new big mosaic has been observed with PMAS/PPak. Its analysis is still in progress (see preliminary results in [36]).

The much better quality of this new mosaic is definitively proven by the detection of the faint C II and O II RLs almost in the whole Huygens region (see **Figure 8**). Density and temperature maps have been obtained from different diagnostic ratios such as [S II] $\lambda6717/\lambda6731$, [Cl III] $\lambda5517/\lambda5537$, or [O III] $\lambda5007/\lambda4363$ (see [36]). The dataset is very valuable to investigate the AD problem as an integrated property and compare the results with what is observed in extragalactic H II regions. It will also be possible to address a comparison with the previous IFU studies and investigate whether the AD problem is subject to dilution effects by using larger apertures. Other research interests such as the validity of the case B recombination theory, the effect of scattered light, the global ionization structure of the nebula, the effect of collisional deexcitation by the presence of high-density morphologies, the accuracy of the RL and CEL methods, or the correlation between physical conditions and chemical abundances with morphologies at large and intermediate scales will be explored thanks to this mosaic and the clear potential of IFS studies.

## 7. Final Remarks: A Bright Future for IFS Studies

The Orion Nebula is a landmark object of the solar neighborhood and serves as a paradigm for interpreting results throughout the Galaxy and beyond. It is a fundamental reference for our knowledge about formation and evolution processes of stars and planets and for evaluating the reliability of the methods used to ascertain the chemical composition in the universe. If H II regions like the Orion Nebula are more the rule rather than an exception, then this nebula is the only target that gives us the opportunity to investigate and understand underlying mechanisms that operate in the interior of these gaseous clouds at small spatial scales. Without any doubt, IFS plays an active role in this framework and exciting results have been, and will be surely, discovered by using this technique.

Applying IFS to unveil the intimate properties of the Orion Nebula is today in its rising stage. At the time that this review was written, a total of 9 IFS studies were found in the literature devoted to the Orion Nebula. As it has been shown along the review, these studies have used the state-of-the-art optical spectrographs PMAS, GMOS, and FLAMES, currently working from ground-based observatories at the 3.5 m telescope of Calar Alto Observatory, the 8 m Gemini telescopes, and the 8.2 m VLT, respectively. The capabilities of IFS have allowed us to enhance our knowledge of physical, chemical, and structural properties of the Orion Nebula and its morphological substructures. For instance, the study of stratified bars has contributed to constrain with more detail the 3D picture of the Orion-S star-forming cloud. The IFS has allowed us to spatially locate and quantify the effects of high-velocity flows on their surrounding media as well as to investigate for the first time the chemical content of proplyds. Furthermore, it has been shown that small-spatial-scale morphologies may be playing an active role in the production of the AD problem. Results from the improved mosaic of Orion will be very useful to evaluate how the small-scale structure affects to the global properties of the nebula, which may give us new clues to understand the AD problem at larger scales.

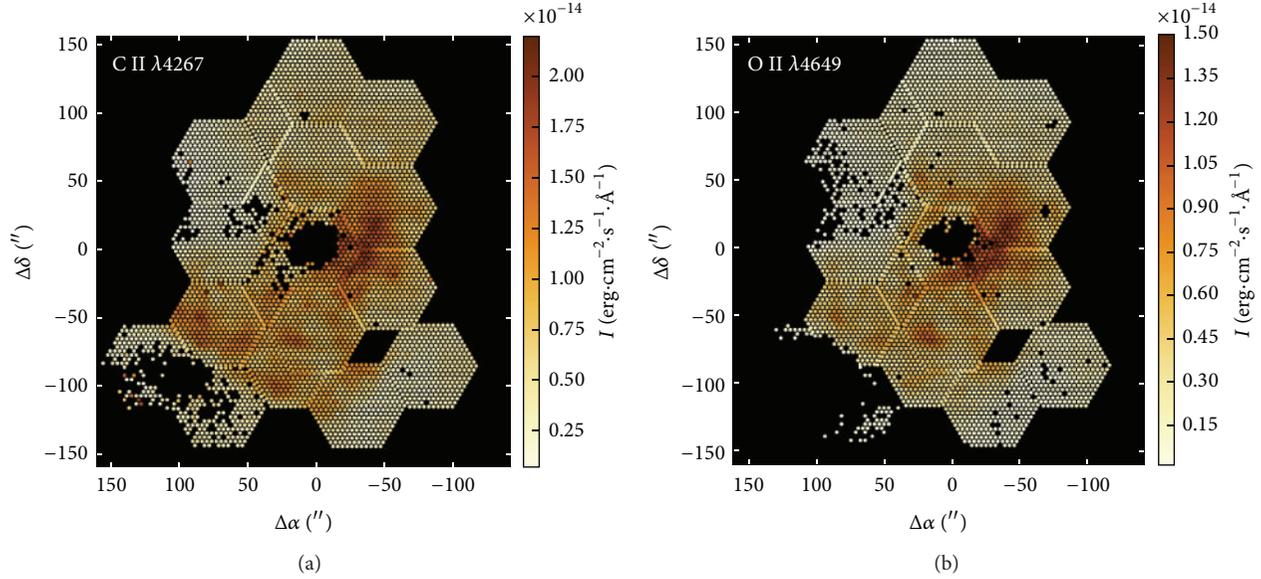

FIGURE 8: Spatial distribution maps of dereddened fluxes of RLs: (a) C II 4267 Å and (b) O II 4649 Å. The black areas in the center are contaminated by emission from the Trapezium stars. The coordinates are centered on $\theta^1$Ori C. The figure has been adapted from [36].

Further investigation is still required in many of the achievements presented here. Second and third generations of integral field spectrographs with improved performances in the optical range will have much to say about it. Of these new generations, it should be highlighted: the simultaneous observation of the near-ultraviolet, optical and near-infrared spectral ranges offered by X-shooter, operating at intermediate resolutions [90]; the large fields of view that will be provided by MUSE [91] or VIRUS [92]; or the development of new IFS techniques as SAMI [93], which mixes the multiobject technique with spatial capability of the IFS. For the future 30 m telescopes, new advances in this area aims to combine the integral field technique with high spectral resolution and adaptive optics systems (even in the optical range), auguring a bright future for the maturity of the IFS and, certainly, for expanding our current knowledge of the Orion Nebula.

## Conflict of Interests

The authors declare that there is no conflict of interests regarding the publication of this paper.

## Acknowledgments

The author is thankful to C. Esteban, W. J. Henney, C. R. O'Dell, T. H. Puzia, and the referee for the revision of the paper as well as for their suggestions, which have improved the readability of this work. The author deeply grateful for the inestimable dedication, support, and contribution of their coworkers: C. Esteban, N. Flores-Fajardo, J. García-Rojas, W. J. Henney, L. López-Martín, V. Luridiana, M. Núñez-Díaz, and Y. G. Tsamis. Without them, much of the reviewed work would not have been possible. The author thanks the organizers of the monographic "Metals in 3D: A Cosmic View from Integral Field Spectroscopy" for the opportunity of being part of it. Finally, the author acknowledges the funding support from Comité Mixto ESO-Chile, the Basal-CATA Grant no. PFB-06/2007 and the FONDECYT project no. 3140383.

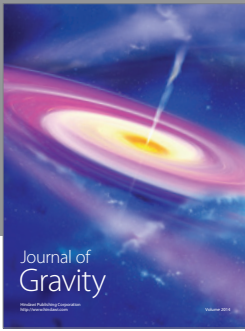
Journal of
Gravity

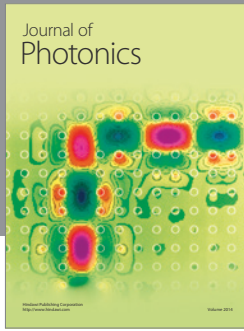
Journal of
Photonics

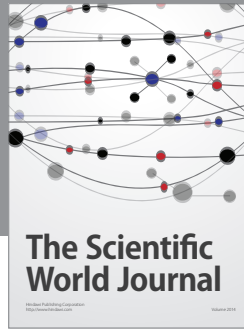
The Scientific
World Journal

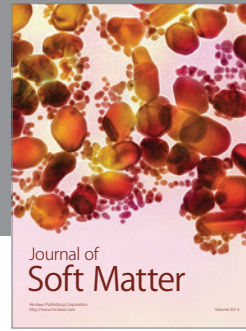
Journal of
Soft Matter

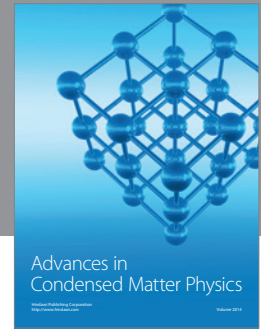
Advances in
Condensed Matter Physics

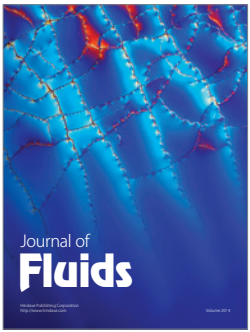
Journal of
Fluids

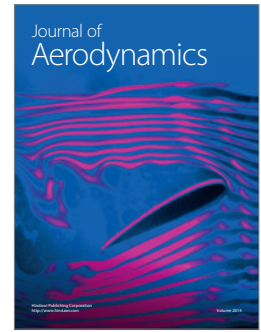
Journal of
Aerodynamics

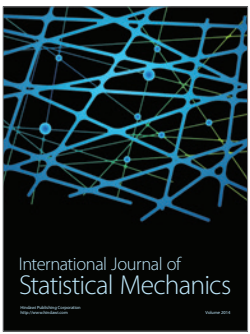
International Journal of
Statistical Mechanics

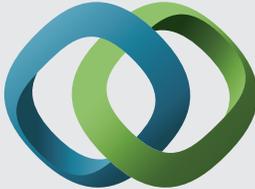

Hindawi

Submit your manuscripts at
http://www.hindawi.com

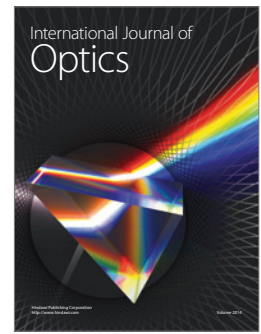
International Journal of
Optics

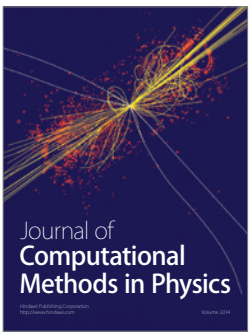
Journal of
Computational
Methods in Physics

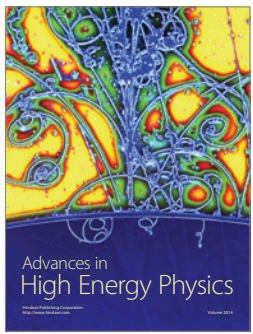
Advances in
High Energy Physics

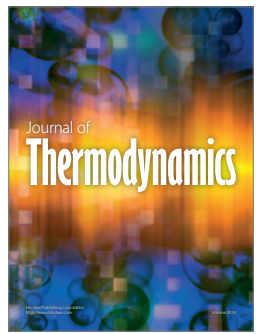
Journal of
Thermodynamics

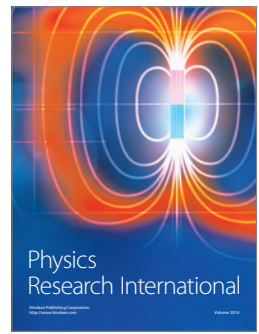
Physics
Research International

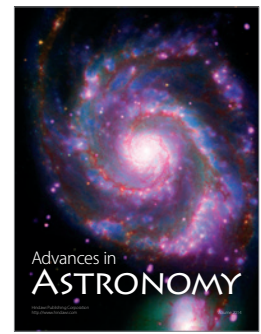
Advances in
ASTRONOMY

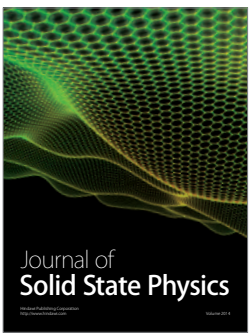
Journal of
Solid State Physics

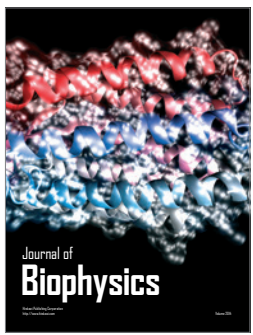
Journal of
Biophysics

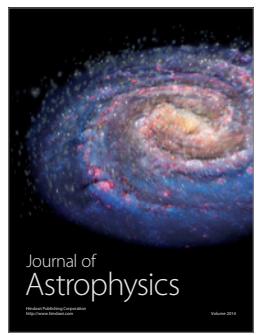
Journal of
Astrophysics

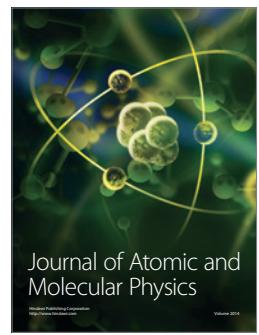
Journal of Atomic and
Molecular Physics

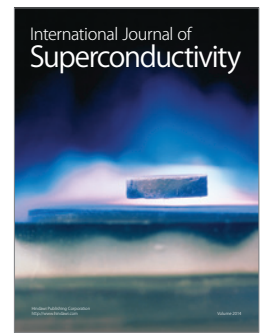
International Journal of
Superconductivity